\documentclass[floatfix,reprint,amsmath,amssymb,aps,longbibliography]{revtex4-2}

\usepackage{float}
\usepackage{graphicx}
\usepackage{dcolumn}
\usepackage{bm}

\usepackage{verbatim}
\usepackage{color,soul}

\usepackage{hyperref}
\hypersetup{
    colorlinks=true,
    linkcolor=blue,
    filecolor=blue,      
    urlcolor=blue,
    citecolor=blue,
}
\begin{document}

\preprint{APS/123-QED}

\title{Force approach for the pseudopotential lattice Boltzmann method}

\author{L. E. Czelusniak\textsuperscript{1}}

\email{luiz.czelusniak@usp.br}

\author{V. P. Mapelli\textsuperscript{1}}

\author{M. S. Guzella\textsuperscript{2}}

\author{L. Cabezas-G\'omez\textsuperscript{1}}

\author{Alexander J. Wagner\textsuperscript{3}}

\affiliation{
\textsuperscript{1}Heat Transfer Research Group,
Department of Mechanical Engineering, S\~ao Carlos School of Engineering,
University of S\~ao Paulo, S\~ao Carlos, SP, Brazil\\
\textsuperscript{2}Institute of Science and Technology,
Federal University of Jequitinhonha and Mucur\'i Valleys, UFVJM,
Diamantina, MG, Brazil\\
\textsuperscript{3}Department of Physics, North Dakota State University, Fargo, North Dakota 58108, USA
}%


\date{\today}

\begin{abstract}
The pseudopotential method is one of the most popular extensions of the lattice Boltzmann method (LBM) for phase change and multiphase flow simulation.
One attractive feature of the original proposed method consists on its simplicity of adding a force dependent on a nearest-neighbor potential function, which became known as the Shan-Chen interaction force.
Some of the well known drawbacks implied by this method involves lack of thermodynamic consistency and impossibility to control the surface tension independently.
In order to correct these deficiencies, different approaches were developed in the literature, such as multirange interactions potential, which involves larger stencils than nearest-neighbor approach, and modified forcing schemes.
In this work, a strategy is developed to control the liquid-gas density ratio and the surface tension by means of an appropriate interaction force field using only nearest-neighbor interactions. The proposed procedure is devised starting from the desired pressure tensor, which allow for the control of the equilibrium multiphase properties such as liquid-gas coexistence curve and surface tension. Then, it is shown how to derive an external force field able to replicate the effects of this pressure tensor in the macroscopic conservation equations. The final step of our procedure is implementing this external force in the LBE by using the classical Guo forcing scheme. Numerical tests regarding static and dynamic flow conditions were performed. Results obtained from simulations showed good agreement with expected analytical values. Most divergent solution observed was the droplet oscillation period under certain flow conditions, which deviated 9\% from expected analytical result. The observed results corroborate that the proposed method is able to replicate the desired macroscopic multiphase behaviour. 
\end{abstract}

\maketitle


\section{Introduction} 
\label{sec:Intro}

The lattice Boltzmann method (LBM) \cite{chen1998lattice} has grown as an alternative tool for fluid flow simulation. 
Differently from other numerical methods based on a direct discretization of the
conservation equations, the LBM is based on a discretized form of the Boltzmann transport equation
known as the lattice Boltzmann equation (LBE) \cite{shan1998discretization}.
In particular, for phase change phenomena and multiphase flow simulation, several models 
were developed within the LBM framework \cite{gunstensen1991lattice,swift1996lattice,shan1993lattice,luo1998unified}. 
One of the most popular is the pseudopotential method \cite{shan1993lattice,shan1994simulation}.
It consists in the definition of an artificial interaction potential which is capable of 
inducing phase separation. In this way, it is not necessary to track the interface between 
multiple phases as they are maintained by the short-range attraction force imposed to the fluid. 
This type of procedure is called diffuse interface modeling \cite{anderson1998diffuse}, since the density field varies
continuously between the different phases due to the action of the force field, instead of 
having an exact interface location. 

The original pseudopotential method was developed by \citet{shan1993lattice}. 
The authors proposed an interaction force that could maintain different phases in equilibrium.
The drawbacks of this procedure involves lack of thermodynamic consistency and
non-adjustable surface tension. 
In a subsequent work, \citet{shan1994simulation} focused on the macroscopic behavior of their
method. The authors addressed the effects of the proposed interaction force into the pressure tensor. 
With this approach, the authors were able to study the equilibrium properties of a fluid governed by this resulting pressure tensor. It is known that
in diffuse interface models, the pressure tensor plays a key role in the phase-change process, 
controlling liquid-gas density ratio and surface tension \cite{li2016lattice}. 
A different interaction force was proposed by \citet{zhang2003lattice}, but
this model suffers from the same issues of the \citeauthor{shan1993lattice} approach. 
The first improvement was done 
by \citet{kupershtokh2009equations}, who were able to adjust the liquid-gas coexistence curve by combining the previous interaction forces. However, 
this technique was still not able to allow controlling surface tension without affecting the liquid-gas densities. 
A similar procedure was also proposed later \cite{gong2012numerical}.
This technique allowed successful applications of LBM to multiphase simulations,
such as simulations of pool boiling \cite{gong2017direct,ma20193d}.

The procedures aforementioned are classified as nearest neighbor
interaction forces, since their implementation requires only information from the fluid properties at
the nodes adjacent to the node of interest. One of the further attempts to enhance multiphase behavior 
consists in the multirange interaction forces \cite{shan2006analysis}, which use larger numerical stencils 
involving nodes at greater distances. 
\citet{sbragaglia2007generalized} proposed a multirange model capable of adjusting the 
liquid-gas density curve and the surface tension. However, \citet{li2013achieving}
noticed that this model had some issues, since the density ratio of the system varied considerably
with the change in surface tension. Recently, \citet{kharmiani2019alternative} proposed
a consistent interaction potential that permits to control independently the liquid-gas 
density ratio and surface tension. But one of the terms that constitutes the proposed
force is calculated in two steps and it can be argued that this procedure is equivalent to a multirange 
approach, since it requires information from distances greater than the adjacent nodes.
The disadvantages of the multirange model involve being 
computationally more expensive and the boundary conditions need to be modified \cite{kruger2017lattice}.
Besides that,
considering a first principles approach mapping a Molecular Dynamics simulation onto the lattice Boltzmann framework \cite{parsa2017lattice} we will argue below, 
that interactions should only involve adjacent nodes in the vast majority of practical simulations. 

In order to incorporate the effects of an external force field into the LBE,
no matter if it is a nearest-neighbour or multirange approach,
one may use numerical procedures known as forcing schemes. 
Very common examples from literature are the forcing schemes developed by \citet{guo2002discrete}, 
\citet{shan1993lattice}, \citet{kupershtokh2004new} and \citet{wagner2006thermodynamic}. 
The use of a suitable forcing scheme in a numerical solution has been shown to have great importance, 
since some authors have observed distinguished 
behaviors for different schemes, even when the same external force field was applied 
\cite{li2012forcing,huang2011forcing}.
\citet{li2012forcing} identified that such distinct behaviors were caused by
distinguished terms introduced into the pressure tensor by the forcing schemes, and that
affected the multiphase properties of the method.
Based on this finding, the authors proposed a source term for the LBE in order to
change the pressure tensor and to control the liquid-gas coexistence curve of the pseudopotential method. 
Later, the procedure was extended to allow the surface tension control without affecting the liquid 
and vapor densities \cite{li2013achieving}. This procedure is very attractive because the numercial scheme
involves only properties at the adjacent nodes, resulting in a computationally efficient method.
Most subsequent approaches in the literature followed this reasoning  
\cite{lycett2015improved,huang2016third,zhai2017pseudopotential}.  
Also, it was discovered that higher order discretization errors caused by the forcing schemes
play a big role in multiphase flows \cite{wagner2006thermodynamic,lycett2015improved} and these errors must be taken into
account for proper determination of the pressure tensor.
These procedures based on the work of \citet{li2012forcing} allowed many applications of the pseudopotential method
\cite{li2015lattice,li2018enhancement,hu20192d}.

Even though many theoretical developments in forcing schemes were achieved concerning the design of
pressure tensors that allow the control of the desired equilibrium multiphase properties,
this knowledge was still not properly employed to devise interaction forces to overcome the limitation of previous models  \cite{shan1993lattice,zhang2003lattice,kupershtokh2009equations}. 
Some attempts were done but they involve the use of multirange interactions which reduce the method computational efficiency.
Based on the current developments in the pseudopotential literature, in this work, we developed a strategy to control the liquid-gas density ratio and the surface 
tension by means of an appropriate interaction force field using only nearest-neighbor interactions, without resorting to a change in the forcing scheme. The procedure starts by considering the desired pressure tensor, which allows for the control of the equilibrium properties of the pseudopotential method. We then derive an external force field which replicates the effects of this pressure tensor in the momentum conservation equation. The final step of our procedure is implementing 
this external force in the LB method by using the classical forcing scheme developed by \citet{guo2002discrete}.

The present paper is organized as follows. In Sec.~\ref{sec:TB}, the theoretical background related to LBM and pseudopotential method will be briefly discussed, with particular focus on the pressure tensor role.
In Sec.~\ref{sec:GIPFP}, a fundamental approach to analyze the form of the interaction force will be discussed. This analysis is used as a foundation for the argument that using adjacent nodes in the pseudopotential method suffices to practical simulations.
Then, in Sec.~\ref{sec:MOTED}, it will be shown how to discretize the terms of the desired pressure tensor using finite differences. Later, an interaction force will be devised to replicate the effect of the desired pressure tensor in the conservation equations as shown in Sec.~\ref{sec:FA}. Numerical simulations will be presented in Sec.~\ref{sec:NS} to validate the proposed interaction force. Finally, a brief conclusion drawn from theoretical and numerical studies will be made in Sec.~\ref{sec:Conclusion}.



\section{Theoretical Background} 
\label{sec:TB}


\subsection{The Lattice Boltzmann Equation} \label{sec:TLBE}

The LBE can be written as:
\begin{equation} \label{eq:TLBE}
f_i(t+1,\bm{x}+\bm{c}_i) - f_i(t,\bm{x}) = \Omega_i(\bm{f},\bm{f}^{eq}) + S_i,
\end{equation}		
where $f_i$ are the particle distribution functions related with the velocity $\bm{c}_i$
and $\bm{f}$ is a vector with components $[\bm{f}]_i=f_i$. 
Also, $t$ and $\bm{x}$ are the time and space coordinates, respectively. 
The term $\Omega_i(\bm{f},\bm{f}^{eq})$ is the collision operator and it is, in general, dependent 
on $\bm{f}$ and the equilibrium distribution function, $\bm{f}^{eq}$. 
For the two-dimensional nine velocities set (D2Q9), the velocities $\bm{c}_i$
are given by:
\begin{equation} 
\label{eq:VelocitySet}
\bm{c}_i =
\begin{cases} 
      (0,0), ~~~~~~~~~~~~~~~~~~~~~~~~~~~~~~~~~~ i = 0,  \\
      (1,0), (0,1), (-1,0), (0,-1), ~~~~~ i = 1,...,4, \\
      (1,1), (-1,1), (-1,-1), (1,-1), ~ i = 5,...,8. \\
   \end{cases}
\end{equation}		

The simplest form of $\Omega_i(\bm{f},\bm{f}^{eq})$ is the single-relaxation time, 
also known as BGK collision operator \cite{bhatnagar1954model}, described in Eq.~(\ref{eq:BGKCO}). 
One can improve stability and to some extent accuracy by allowing different relaxation times for different modes. This is known as the  
multi-relaxation time (MRT) collision operator, shown in Eq.~(\ref{eq:MRTCO}).
\begin{subequations}
\begin{equation} 
\label{eq:BGKCO}
\Omega_i(\bm{f},\bm{f}^{eq}) = - \frac{1}{\tau} (f_i - f_i^{eq}),
\end{equation}	
\begin{equation} 
\label{eq:MRTCO}
\Omega_i(\bm{f},\bm{f}^{eq}) = - \left[ \bm{M}^{-1} \bm{\Lambda} \bm{M} \right]_{ij}  (f_j - f_j^{eq}),
\end{equation}	
\end{subequations}
where the parameter $\tau$, in Eq.~(\ref{eq:BGKCO}), is the relaxation time. In Eq.~(\ref{eq:MRTCO}), 
$\bm{\Lambda}$ is the relaxation matrix and $\bm{M}$ is the matrix that converts $(\bm{f} - \bm{f}^{eq})$ 
into a set of moments. The particular form of these matrices can vary, as discussed by \citet{kaehler2013derivation}, but the hydrodynamic modes of mass, momentum, and stress tensor have to be eigenvectors of the collision matrix. The eigenvalues of this matrix then represent now a set of relaxation times that can be different for the different eigenvectors. 
The MRT collision operator has been widely used in multi-phase simulations \cite{li2013lattice,li2015lattice,mu2017nucleate}. Note that the MRT equation recovers the BGK collision operator if all relaxation times of the MRT collision operator are equal.
The form of the matrices $\bm{M}$ and $\bm{\Lambda}$ are presented in 
Appendix~\ref{sec:MRT}.

A popular form of the equilibrium distribution function is:
\begin{equation} 
\label{eq:SEDF}
f_i^{eq} = w_i \bigg( \rho + \frac{c_{i \alpha}}{c_s^2} \rho u_{\alpha}
+ \frac{(c_{i \alpha}c_{i \beta} - c_s^2 \delta_{\alpha \beta})}{2 c_s^4} 
\rho u_{\alpha}u_{\beta} \bigg),
\end{equation}	
where the terms $w_i$ are the weights related with each velocity $\bm{c}_i$, and $c_s$ is
the lattice sound speed. For D2Q9 set, the weights $w_i$
are given by $w_0 = 4/9$, $w_{1,2,3,4} = 1/9$ and $w_{5,6,7,8} = 1/36$, and $c_s$ is equal $1/\sqrt{3}$.
Also, $\rho$ and $\bm{u}$ are the fluid density and velocity, respectively given by (\ref{eq:FD}) and (\ref{eq:FMD}).

The last term in the right-hand side of Eq.~(\ref{eq:TLBE}), $S_i$, is what defines the forcing scheme, 
i.e. this term is responsible for adding the effects of an external force field, $F_{\alpha}$, 
in the recovered macroscopic conservation equations. 
One of the most widely used forcing scheme in literature was developed by \citet{guo2002discrete},
and it can be described as follows: 
\begin{equation} 
\label{eq:GFS}
S_i  = C_{ij} w_j \bigg( \frac{c_{j \alpha}}{c_s^2} F_{\alpha}
+ \frac{(c_{j \alpha}c_{j \beta} - c_s^2 \delta_{\alpha \beta})}{c_s^4} 
F_{\alpha}u_{\beta} \bigg),
\end{equation}	
where the term $C_{ij}$ depends on whether the BGK, Eq.~(\ref{eq:GFSBGK}), or the MRT, 
Eq.~(\ref{eq:GFSMRT}), collision operator is being used. Both definitions can, respectively,
be given by:
\begin{subequations}
\begin{equation} 
\label{eq:GFSBGK}
C_{ij}  = \bigg( 1 - \frac{1}{2 \tau} \bigg) \delta_{ij},
\end{equation}	
\begin{equation} 
\label{eq:GFSMRT}
C_{ij}  = \left[ \bm{M}^{-1} \bigg( \bm{I} - \frac{\bm{\Lambda}}{2} \bigg) \bm{M} \right]_{ij},
\end{equation}	
\end{subequations}
where $\bm{I}$ is the identity matrix. The relation between particle distribution 
functions $f_i$ and the actual fluid velocity $\bm{u}$ depends on the forcing scheme.
For the \citeauthor{guo2002discrete} forcing scheme, density and velocity 
fields are given by:
\begin{subequations}
\begin{equation} 
\label{eq:FD}
\rho = \sum_i f_i,
\end{equation}	
\begin{equation} 
\label{eq:FMD}
\rho \bm{u} = \sum_i f_i \bm{c}_i + \frac{\bm{F}}{2}.
\end{equation}	
\end{subequations}

The momentum density shown in Eq.~(\ref{eq:FMD}) needs to take into account the force field
term, $\bm{F}/2$, in order for the numerical scheme to recover second-order
accurate conservation equations under the influence of an external force field.

The LBE describes the evolution of particle distribution functions, however, the variables 
of interest are the macroscopic flow fields. The correspondence between the LBE and the
macroscopic behavior that it simulates can be shown through different approaches. 
The standard procedure is the Chapman-Enskog analysis, and one alternative is the 
recursive substitution developed by \citet{wagner1997theory} and further developed by \citet{holdych2004truncation} and \citet{kaehler2013derivation}. 
Up to second order terms, both procedures result in the same behavior, and it is not known if differences at higher orders will occur. 
Either approach recovers the mass and momentum conservation equations to second order:
\begin{subequations}
\begin{equation} 
\label{eq:MCE}
\partial_t \rho + \partial_{\alpha} (\rho u_{\alpha}) = 0,
\end{equation}	
\begin{equation} 
\label{eq:NSMCE}
\partial_t (\rho u_{\alpha}) + \partial_{\beta} (\rho u_{\alpha} u_{\beta}) =
- \partial_{\beta} p_{\alpha \beta}
+ \partial_{\beta} \tau_{\alpha \beta} + F_{\alpha},
\end{equation}	
\end{subequations}
where the stress tensor, $\tau_{\alpha \beta}$, is given by 
$\tau_{\alpha \beta} = \rho c_s^2 (\tau - 0.5) (\partial_{\beta} u_{\alpha} + \partial_{\alpha} u_{\beta})$ 
for the BGK collision operator, Eq.~(\ref{eq:BGKCO}). The pressure tensor is given by 
$p_{\alpha \beta}=\rho c_s^2 \delta_{\alpha \beta}$. When MRT collision operator is applied, 
it is possible to adjust the bulk and shear viscosity independently in the stress tensor,
since a greater number of relaxation times are used.
A more thorough analysis applying the MRT collision operator can be seen in the work
of \citet{kaehler2013derivation}.  

Even though the LBE recovers the correct form of the Navier-Stokes up to second order terms,
several studies have shown that the third order spatial discretization errors due to the 
forcing scheme play an important role in pseudopotential methods. These errors must be 
taken in account for the correct multiphase behavior prediction of the method. 
Third order analysis of the LBE considering different forcing schemes have been carried out in the LB literature. 
\citet{zhai2017pseudopotential}, through Chapman-Enskog analysis, evaluated
the recovered macroscopic equations up to the third order, considering the \citeauthor{guo2002discrete} forcing scheme. 
\citet{lycett2015improved} also investigated third order therms of a generic
forcing scheme, using the technique developed by \citeauthor{holdych2004truncation}. 
From the results of these studies, one can show that third order discretization error produced
by the \citeauthor{guo2002discrete} forcing scheme is given by:
\begin{equation} 
\label{eq:TOSDEFS}
E_{\alpha}^{3rd} = \frac{c_s^2}{12} \partial_{\beta} \Big[ 
(\partial_{\gamma} F_{\gamma}) \delta_{\alpha \beta} 
+ \partial_{\alpha} F_{\beta} + \partial_{\beta} F_{\alpha} \Big],
\end{equation}	
this term should be added in the right-hand side of Eq.~(\ref{eq:NSMCE}) in 
order to take into account the influence of higher order error in pseudopotential methods.



\subsection{Pressure Tensor and Phase Change} \label{sec:PMPC}

A common approach to address to multiphase lattice Boltzmann simulations is to define the force
field to be implemented in LBE, and then to analyze the resulting pressure tensor, from
which it is possible to draw conclusions of key multiphase features, such as equation of
state, liquid-gas coexistence curve and surface tension.

In this work, we use a general pressure tensor as starting point, and show how it is 
related to multiphase flow properties. Afterwards, in next sessions, it is shown how it
can be implemented through a discrete force in LBE, and how it is possible to devise a better method
when compared to original Shan-Chen formulation.  

A general pressure tensor from a single-phase pseudopotential method can be written as:
\begin{equation} 
\label{eq:GPTEF}
\begin{aligned}
p_{\alpha \beta} &= \left(  c_s^2 \rho + G \psi^2 + C_1 G (\partial_\gamma \psi)(\partial_\gamma \psi)
+ C_2 G \psi \partial_{\gamma} \partial_{\gamma} \psi \right) \delta_{\alpha \beta} \\ 
&+ C_3 G (\partial_{\alpha} \psi) (\partial_{\beta} \psi)	
+ C_4 G \psi \partial_{\alpha}\partial_{\beta} \psi,
\end{aligned}
\end{equation}
where $C_{1,2,3,4}$ are arbitrary coefficients, $\psi$ is a density-dependent interaction
potential, $\psi=\psi(\rho)$, and $G$ is a parameter that controls the strength of interaction.
One should notice that for a uniform state, the pressure tensor is simplified to 
$p_{\alpha \beta}=\left(  c_s^2 \rho + G \psi^2 \right) \delta_{\alpha \beta}$.
This term plays the role of the equation of state, and upon this fact, \citet{yuan2006equations}
proposed the following definition: 
\begin{equation} 
\label{eq:DED}  
\psi = \sqrt{\frac{P_{EOS}-c_s^2 \rho}{G}},
\end{equation}	
where the term $P_{EOS}$ represents any desired equation of state to be introduced into the method.
When this technique is used, parameter $G$ no longer controls the interaction strength,
and it can be seen as an auxiliary parameter to keep the term inside the square root 
positive.

Observing the recovered momentum conservation equation in Eq.~(\ref{eq:NSMCE}), one 
may notice that what affects momentum balance is the divergence of the pressure tensor, 
$-\partial_{\beta}p_{\alpha \beta}$, and not the pressure tensor itself.
Therefore, as pointed out by \citet{sbragaglia2007generalized},
different pressure tensors can reproduce identical hydrodynamic behaviors, as long as their 
divergences are equal to each other. 

By applying the following tensor identity (for more details refer to Appendix~\ref{sec:TI}):
\begin{eqnarray} 
\label{eq:TI} 
\partial_{\beta} \big[ \psi \partial_{\alpha} \partial_{\beta} \psi - \big( \psi \partial_{\gamma} \partial_{\gamma} \psi \big) \delta_{\alpha \beta} \big] = &&~
\partial_{\beta} \big [ (\partial_{\gamma} \psi) (\partial_{\gamma} \psi) \delta_{\alpha \beta} \nonumber \\
&& - (\partial_{\alpha} \psi) (\partial_{\beta} \psi) \big],				
\end{eqnarray} 	 
it is possible to show that the divergence of the tensor given by
Eq.~(\ref{eq:GPTEF}) is equivalent to the divergence of the following pressure tensor:
\begin{equation} 
\label{eq:GPTRF}
\begin{aligned}
p_{\alpha \beta} &= \left(  c_s^2 \rho + G \psi^2 + A_1 G (\partial_\gamma \psi)(\partial_\gamma \psi)
+ A_2 G \psi \partial_{\gamma} \partial_{\gamma} \psi \right) \delta_{\alpha \beta} \\ 
&+ A_3 G \psi \partial_{\alpha}\partial_{\beta} \psi,	
\end{aligned}
\end{equation}
where $A_{1,2,3}$ are arbitrary coefficients that obey the following relations:
$A_1=C_1+C_3$, $A_2=C_2+C_3$ and $A_3=C_4-C_3$. 
The reduced form of pressure tensor, Eq.~(\ref{eq:GPTRF}), is going to be used
throughout the text. 

A suitable problem to check liquid-gas coexistence curve and thermodynamic consistency obtained 
from the pressure tensor presented before is the planar interface between two phases in
mechanical equilibrium \cite{shan2008pressure}. Assuming $x$ and $y$ as the coordinates in,
respectively, the normal and tangential direction to the interface, one may simplify the pressure
tensor, once there is no gradients in $y$-direction, to: 
\begin{subequations}
\begin{equation}
\label{eq:PTNNPI} 
p_{x x} = c_s^2 \rho + G \psi^{2} 
+ G \Big[ A_1 \Big( \frac{d \psi}{dx} \Big)^{2}
+ (A_2+A_3) \psi \frac{d^{2} \psi}{dx^{2}} \Big],	
\end{equation}
\begin{equation} 
\label{eq:PTTTPI} 
p_{y y} = c_s^2 \rho + G \psi^{2} 
+ G \Big[ A_1 \Big( \frac{d \psi}{dx} \Big)^{2}
+ A_2 \psi \frac{d^{2} \psi}{dx^{2}} \Big],	
\end{equation}	
\begin{equation} 
\label{eq:PTNTPI}  
p_{x y} = p_{y x} = 0.	
\end{equation}	
\end{subequations}

The mechanical equilibrium condition implies that the pressure tensor component $p_{xx}$ must be 
constant and equal to the bulk pressure $p_{0}$ along the $x$ axis. By imposing this condition, 
\citet{shan2008pressure} deduced that the gas and liquid densities obtained by the pseudopotential 
method must satisfy the following relation: 
\begin{equation} 
\label{eq:PGLDR}
\int_{ \rho_{g} }^{ \rho_{l} } \left( p_{0} - c_s^2 \rho - G \psi^{2} \right)	\frac{\dot{\psi}}{\psi^{1+\epsilon}} d \rho = 0,
\end{equation}	
where $\epsilon=-2A_1/(A_2+A_3)$ and $\rho_{l}$, $\rho_{g}$ are the densities of the liquid and vapor phases, respectively. The dot, as in $\dot{\psi}$, denotes the derivative with respect to density $\rho$.
Another consequence of the equilibrium condition is that the bulk pressure of liquid and
vapor regions far away from the interface must also be equal to $p_0$:
\begin{subequations}
\begin{equation} 
\label{eq:BPLP}
p_{0} = c_s^2 \rho_{l} + G [\psi(\rho_{l})]^{2},
\end{equation}
\begin{equation} 
\label{eq:BPVP}
p_{0} = c_s^2 \rho_{g} + G [\psi(\rho_{g})]^{2}.
\end{equation}
\end{subequations}

Together, Eqs.~(\ref{eq:PGLDR}), (\ref{eq:BPLP}) and (\ref{eq:BPVP}) compose a
well-posed problem that can be solved for $p_{0}$, $\rho_{l}$ and $\rho_{g}$.
In fact, this problem resembles the Maxwell equal-area rule, which states that for
a given temperature, a thermodynamic consistent phase-change obeys the 
following gas-liquid density relation:
\begin{equation} 
\label{eq:MLVDR}
\int_{ \rho_{g} }^{ \rho_{l} } \left( p_{0} - P_{EOS} \right)	\frac{d \rho}{\rho^{2}}  = 0.
\end{equation}	

By comparing Eqs.~(\ref{eq:MLVDR}) and (\ref{eq:PGLDR}), \citeauthor{lycett2015improved}
were able to conclude that correct thermodynamic consistency will be achieved when:
\begin{equation} 
\label{eq:CTC} 
\frac{\dot{\psi}}{\psi^{1+\epsilon}} d \rho \propto \frac{d \rho}{\rho^{2}}.
\end{equation}	

From Eq.~(\ref{eq:CTC}) it is clear that the thermodynamic consistency of the 
pseudopotential method depends on the equation of state used to define the interaction potential
and on parameter $\epsilon$, which in turn, are related to coefficients of pressure tensor.

Another important aspect of multiphase simulation is to properly control the surface tension. 
According to \citet{rowlinson2013molecular}, the surface tension in diffuse interface models 
can be defined as:
\begin{equation} 
\label{eq:STDIM}
\gamma = \int_{-\infty}^{\infty} \Big( p_{xx} - p_{yy} \Big) dx,
\end{equation}	
where, again, $x$ and $y$ are the interface normal and tangential directions, respectively.
Equation (\ref{eq:STDIM}) implies that the surface tension depends only on the anisotropic 
part of the pressure tensor.
And, by consequence, it can be adjusted by the parameter $A_3$. 
For a planar interface, Eqs.~(\ref{eq:PTNNPI}) and (\ref{eq:PTTTPI}) must be inserted into Eq.~(\ref{eq:STDIM}), resulting in the following relation:
\begin{equation} 
\label{eq:STPI}
\gamma_{pi} = \int_{-\infty}^{\infty} A_3 \psi \frac{d^{2} \psi}{dx^{2}} dx,
\end{equation}	

In order to compute the surface tension of the planar interface case $\gamma_{pi}$,
one can obtain the density profile that solve Eq.~(\ref{eq:PTNNPI}) (for specific values of the parameters $A_1$, $A_2$ and $A_3$) using a numerical method.
This differential equation can be solved replacing the derivatives
by finite difference approximations,
as for example, second order central
differences. 
The resultant nonlinear set of equations can be solved using Newton-Raphson method with 
the phase densities (obtained by solving Eq.~(\ref{eq:PGLDR})) at the borders
as boundary conditions.
With the knowledge of the density profile $\rho(x)$, the interaction potential profile
is determined $\psi(x)=\psi(\rho(x))$.
After that the surface tension can be computed using a numerical integration procedure
to integrate Eq.~(\ref{eq:STPI}).



\subsection{Shan-Chen method} \label{sec:EBPT}

The pseudopotential method originated when \citet{shan1993lattice} proposed 
a interaction force similar to: 
\begin{equation} 
\label{eq:SCF}
F_\alpha^{SC} = - \psi ( \bm{x} ) \frac{2G}{c_s^2} \sum w_i \psi ( \bm{x} + \bm{c}_i ) c_{i\alpha}.				
\end{equation}

Using Taylor series expansion, a continuum form of the \citeauthor{shan1993lattice} force is obtained:
\begin{equation} 
\label{eq:CFSCF}
F_\alpha^{SC} = - G \Big( \partial_{\alpha} \psi^2 + c_{s}^{2} \psi \partial_{\alpha} \Delta \psi + ... \Big).		
\end{equation}

The momentum conservation equation, Eq.~(\ref{eq:NSMCE}), shows that the natural pressure tensor 
of the LBM is $p_{\alpha \beta}=c_s^2 \rho \delta_{\alpha \beta}$.
Neglecting the higher order terms in Eq.~(\ref{eq:CFSCF}), it is possible to introduce this force 
into the pressure tensor using the relation 
$-\partial_{\beta} p_{\alpha \beta}^{SC}=-\partial_{\alpha} (\rho c_s^2\delta_{\alpha \beta}) + F_{\alpha}^{SC}$.
The following relation is obtained:
\begin{align} 
\label{eq:SCPTWDE}
p_{\alpha \beta}^{SC} = & \bigg( c_s^2 \rho + G \psi^2 - \frac{c_s^2 G}{2} (\partial_\gamma \psi)(\partial_\gamma \psi) \bigg) \delta_{\alpha \beta} \nonumber \\
& + c_s^2 G \psi \partial_{\alpha} \partial_{\beta} \psi.
\end{align}
This pressure tensor does not give the correct results for the coexistence 
curve. It is necessary to take into account the effect of the third order spatial discretization 
errors of the forcing scheme. For the \citeauthor{guo2002discrete} forcing scheme, 
this error is given by Eq.~(\ref{eq:TOSDEFS}). 

It is possible to evaluate the new pressure tensor, by substituting Eq.~(\ref{eq:CFSCF}) 
into Eq.~(\ref{eq:TOSDEFS}). For simplification, here will be considered 
$F_{\alpha}^{SC} \approx -G \partial_{\alpha} \psi^2$. In this way, using Eq.~(\ref{eq:TI}), 
the discretization errors assume the form:
\begin{align}
E_{\alpha}^{3rd} = & - \partial_{\beta} \left( \frac{c_s^2 G}{2} (\partial_\gamma \psi)(\partial_\gamma \psi)
+ \frac{c_s^2 G}{2} \psi \partial_{\gamma} \partial_{\gamma} \psi \right) \delta_{\alpha \beta} \nonumber \\
& = -\partial_{\beta} p_{\alpha \beta}^{3rd},
\end{align}
where $p_{\alpha \beta}^{3rd}$ is the effect caused by the third order discretization errors 
in the pressure tensor. Adding $p_{\alpha \beta}^{3rd}$ to Eq.~(\ref{eq:SCPTWDE}), the correct 
form of the pressure tensor for the pseudopotential method using the \citeauthor{shan1993lattice} force and the 
\citeauthor{guo2002discrete} forcing scheme is obtained: 
\begin{align} 
\label{eq:CFPTSF}
p_{\alpha \beta}^{SC} = & \Big(  c_s^2 \rho + G \psi^2 + \frac{c_s^2 G}{2} \psi \partial_{\gamma} \partial_{\gamma} \psi \Big) \delta_{\alpha \beta}   \nonumber\\
& + c_s^2 G \psi \partial_{\alpha} \partial_{\beta} \psi.		
\end{align}

This highlights a key limitation of the Shan-Chen method: it is not possible to adjust 
the coexistence density curve, dependent on the pressure tensor, and the surface tension independently since they are both derived from the interaction potential $\psi$. 




\section{Incorporating the Pressure Tensor into the LBM} 
\label{sec:DPTPM}

We propose a top-down approach to overcome the limitations inherent  in the Shan-Chen method.
The starting point is the complete pressure tensor, Eq.~(\ref{eq:GPTRF}).
Suitable force fields are devised to add the effect of the desired terms
of this tensor into the recovered macroscopic conservation equations. 
Then, this interaction forces are directly discretized and later they are incorporated in the LBE. 

In Sec.~\ref{sec:GIPFP}, we present a general inter-particle force for the pseudopotential model. 
After that, in Sec.~\ref{sec:MOTED}, we discuss how to obtain numerical approximations to discretize the force terms. 
In Sec.~\ref{sec:FA} we discuss the method used to incorporate
the effect of the desired pressure tensor in the recovered macroscopic conservation equation.


\subsection{Fundamental inter-particle force calculation} 
\label{sec:GIPFP}

Originally the Shan-Chen method was developed with a microscopic interaction picture in mind. We review here an approach to make this relation more direct. A fundamental approach to analyze a lattice Boltzmann method is given by the Molecular Dynamics Lattice Boltzmann (MDLG) approach developed by Parsa \textit{et al.} in \cite{parsa2017lattice}. The key idea here is to map a Molecular Dynamics (MD) simulation onto a lattice gas. The Boltzmann average of this lattice gas is then, in some sense, the most fundamental definition of a lattice Boltzmann method. This approach has proven useful in analyzing the fluctuations in non-ideal systems \cite{parsa2019large}. Here we use a theoretical approach to write down a fundamental representation of the lattice Boltzmann forcing term.

In a MD simulation, the conservative force $\bm{F}$ on one particle is computed considering the 
potential energy $V_{jk}$ between particles $j$ and $k$ by:
\begin{eqnarray} 
\label{eq:CFOP}	
\bm{F}_{j} = - \sum_{k} \partial_{\bm{x}_{k}} V_{jk} ( | \bm{x}_{j} - \bm{x}_{k} | ).
\end{eqnarray}

Formally, we can write this as a continuous force field obtained from an integral over densities:
\begin{eqnarray} 
\label{eq:CFF}	
\bm{F}(\bm{x},t) = \rho(\bm{x},t) \int d\bm{x}' \rho(\bm{x}',t) \partial_{\bm{x}'} V( | \bm{x} - \bm{x}' | ),
\end{eqnarray}
where we define the density as $\rho(\bm{x},t)=\sum_{j} \delta(\bm{x}-\bm{x}_{j}(t))$. The key here is that in 
LB we have lattice cells that receive momentum from their neighboring cells. This is why we now coarse-grain the MD simulation onto a lattice. We define a discrete covering of lattice cells, and a function $\Delta_\zeta(\bm{x})$ which indicates whether the position $x$ is contained in the lattice cell $\zeta$. We then integrate the force
field over a lattice site to give:
\begin{eqnarray} 
\label{eq:FFOLS}
&&~ \tilde{\bm{F}}(\zeta,t) = \int d\bm{x} \bm{F}(\bm{x},t) \Delta_{\zeta} (\bm{x}) \nonumber \\
&& = \int d\bm{x} \rho(\bm{x},t)\Delta_{\zeta} (\bm{x}) \int d\bm{x}' \rho(\bm{x}',t) \partial_{\bm{x}'} V( | \bm{x} - \bm{x}' | ),
\end{eqnarray}
where the lattice space is represented by:
\[
\Delta_{\zeta}(\bm{x}) =
\begin{cases} 
      1,\mbox{if $\bm{x}$ is in } \zeta \\
      0, \mbox{ else} \\
   \end{cases}.
\]

Now in order to consider the interaction between particles from different lattice sites
the last integral is translated into the next sum, which means that the space is now fully decomposed into
lattice sites:
\begin{eqnarray} 
\label{eq:FCIBDLS}
&&~ \int d\bm{x}' \rho(\bm{x}',t) \partial_{\bm{x}'} V( | \bm{x} - \bm{x}' | ) \nonumber \\
&& = \sum_{\eta} \int d\bm{x}' \rho(\bm{x}',t) \Delta_{\eta}(\bm{x}') \partial_{\bm{x}'} V( | \bm{x} - \bm{x}' | ).
\end{eqnarray}

Thus, this sum over $\eta$ is introduced into Eq.~(\ref{eq:FFOLS}) to obtain a force
representation related to the lattice Boltzmann force that means also a sum over 
neighboring lattice sites:
\begin{eqnarray} 
\label{eq:MT}
&&~ \tilde{\bm{F}}(\zeta,t) = \int d\bm{x} \rho(\bm{x})\Delta_{\zeta} (\bm{x}) \nonumber \\
&& \times \sum_{\eta} \int d\bm{x}' \rho(\bm{x}') \Delta_{\eta}(\bm{x}') \partial_{\bm{x}'} V( | \bm{x} - \bm{x}' | ).
\end{eqnarray}

However, this is only an instantaneous force. A lattice Boltzmann (or lattice gas) method has a finite time step $\Delta t$, and the forcing term includes all the momentum absorbed during this finite time-step \cite{li2007symmetric}. The total amount of momentum $\bm{a}$ obtained is then:
\begin{equation}
    \bm{a}(\zeta,T) = \int_{T\Delta t}^{(T+1)\Delta t} \tilde{\bm{F}}(\zeta,t) \; dt,
\end{equation}
where $T$ is the integer time of the simulation. 

This is a fluctuating quantity, since it depends on the microscopic details of initial
particle occupation. The next step is the application of the Boltzmann average, to look
to all possible distributions that are compatible with the given macroscopic state. 
The definition of the probability of finding a particular configuration is then assumed
to follow some local equilibrium assumption.
\begin{eqnarray} 
\label{eq:FFCIBDLS}
\tilde{\bm{a}} = \langle \bm{a}(\zeta,T) \rangle = \int d\rho(\bm{x},t) P(\rho(\bm{x},t)) \bm{a}(\zeta,T).
\label{eqn:a}
\end{eqnarray}

In an isothermal equilibrium system, the probability for a configuration $\rho(\bm{x})$ would be given by:
\begin{eqnarray} 
\label{eq:DP}
P(\rho(\bm{x})) = \frac{1}{Z} e^{ - \frac{H(\rho(\bm{x}))}{k_{B}T} },
\end{eqnarray}
where $H(\rho(\bm{x}))$ is the energy associated with the configuration $\rho(\bm{x})$, $k_B$ is the Boltzmann constant and $T$ the temperature. $Z$ is the partition function.
In a general non-equilibrium situation, however, finding the probability of a density configuration is more challenging. Nonetheless, it is usual to make the assumption of local equilibrium for each lattice site, i.e. assuming that the particle configurations are in (or very close to) a local-equilibrium configuration with the constraint of the coarse-grained lattice densities at different lattice sites, which will be out of equilibrium. 

Analytically deriving a force using Eq.~(\ref{eq:FFCIBDLS}) is a difficult computational task which we leave to a future publication. Here we want to point to a feature that appears when the size of the lattice is much larger than the interparticle interaction range and the mean-square displacement during a timestep $\Delta t$ is likewise much smaller than that a lattice site, as is common in macroscopic and mesoscopic lattice Boltzmann applications: in this case the term $\bm{a}$ only depends on a close neighborhood of lattice sites around the site we are considering. 

We therefore propose, as an Ansatz, a general force for LB as expressed by Eq.~(\ref{eq:Ansatz}) which preserves the locality of the forcing term predicted by (\ref{eq:FFCIBDLS}). 
Like the standard Shan-Chen approach this force contains the interaction potential function $\psi = \psi(\rho)$ which can be adjusted 
to obtain the desired pressure tensor given by Eq.~(\ref{eq:GPTEF}):
\begin{eqnarray} 
\label{eq:Ansatz}
F_{\alpha} = \sum_{i} \sum_{j} A_{ij} \psi(\rho(\bm{x}+\bm{c}_i)) \psi(\rho(\bm{x}+\bm{c}_j)).
\end{eqnarray}

As in the original Shan-Chen approach the function $\psi$ and the tensor $A_{ij}$ are then adjusted, such that we obtain the desired expression of pressure tensor, Eq.~(\ref{eq:GPTEF}). 
It should be noted that, in contrast to the approaches by \cite{sbragaglia2007generalized,kharmiani2019alternative},  the general force field represented by Eq.~(\ref{eq:Ansatz}) 
is formulated considering the nearest-neighbor lattices only. 



\subsection{Interaction potential moments} 
\label{sec:MOTED}

The pressure tensor, Eq.~(\ref{eq:GPTRF}), is composed by the interaction potential function and its 
spatial derivatives. Thus, any attempt to evaluate it shall inevitably involve some numerical approximations for these derivatives. One of the simplest procedures would be using finite difference stencils
to perform these approximations.
A deeper and thorough explanation about these can be found in any classic finite difference method textbook \cite{leveque2007finite}.
By analyzing Eq.~(\ref{eq:SCF}), one may realize that the numerical scheme of the Shan-Chen force
can be interpreted as calculating the discrete first order moment of the term $\psi(\bm{x}+\bm{c}_i)$. 
In this section, a procedure to obtain the finite difference schemes written in the notation
of these moments will be presented.

As only nearest-neighbor interactions are being considered, for the position $\bm{x}$, the 
operations must be done only with the values of the interaction potential $\psi(\bm{x}+\bm{c}_i)$.
The Taylor series expansion of this term is given as follow:
\begin{eqnarray}
\label{eq:TSEED}	
\psi(\bm{x}+\bm{c}_i) = &&~ \psi(\bm{x}) + c_{i \alpha} \partial_{\alpha} \psi(\bm{x})
+ \frac{1}{2} c_{i \alpha}c_{i \beta} \partial_{\alpha} \partial_{\beta} \psi(\bm{x})  \nonumber \\
&& + \frac{1}{6}c_{i \alpha}c_{i \beta}c_{i \gamma} \partial_{\alpha}	\partial_{\beta} \partial_{\gamma} \psi(\bm{x}) + ...
\end{eqnarray}

In Eq.~(\ref{eq:TSEED}), one may observe that in each of the right-hand side terms, there is a 
polynomial in variables related to the lattice velocities.
As for example, the first three terms involves, respectively, $1$, $c_{i \alpha}$ and $c_{i \alpha}c_{i \beta}$. 
Since it is a common practice to represent functions by discrete Hermite expansions in the LBM literature, 
it would be very convenient to rewrite the terms of Eq.~(\ref{eq:TSEED}) in the following form: 
\begin{align} 
\label{eq:HEED}	
w_i \psi(\bm{x}+\bm{c}_i) = & w_i \bigg[ M^0 + \frac{ c_{ i\alpha } }{ c_s^2 } M_\alpha^1 
+ \frac{ c_{i\alpha} c_{i\beta} - c_s^2 \delta_{ \alpha \beta } }{ 2c_s^4 } M_{ \alpha \beta }^2 \nonumber \\
& + ... \bigg].
\end{align}

The moments of $w_i \psi(\bm{x}+\bm{c}_i)$ are given by the following relations:
\begin{subequations}
\begin{equation} 
\label{eq:ZOMED}
M^0 = \sum_i w_i \psi(\bm{x}+\bm{c}_i)  \approx \psi(\bm{x}) + \frac{c_s^2}{2} \Delta \psi(\bm{x}),
\end{equation}
\begin{equation} 
\label{eq:FOMED}
M_{\alpha}^1 = \sum_i w_i c_{i \alpha} \psi(\bm{x}+\bm{c}_i) \approx 
c_s^2 \partial_{\alpha} \psi(\bm{x}) 
+ \frac{c_s^4}{2} \partial_{\alpha} \Delta \psi(\bm{x}),
\end{equation}
\begin{equation} 
\label{eq:SOMED}
M_{\alpha \beta}^2 = \sum_i w_i (c_{i\alpha} c_{i\beta} - c_s^2 \delta_{ \alpha \beta }) \psi(\bm{x}+\bm{c}_i) \approx c_s^4 \partial_{\alpha} \partial_{\beta} \psi(\bm{x}).
\end{equation}
\end{subequations}

It is worth mentioning that, when using the D2Q9 lattice, there are nine linearly independent discrete Hermite polynomials. 
In Eq.~(\ref{eq:HEED}), only six of them were used, since they suffice for the purposes of the present work.



\subsection{Developing a force approach} 
\label{sec:FA}

As discussed in section~(\ref{sec:PMPC}), the terms of the pressure tensor controls the 
multi-phase properties of the pseudopotential method. In particular, the following terms
are useful: 
\begin{subequations}
\begin{equation} 
\label{eq:PTCCC}
p^{(1)}_{\alpha \beta} = (\partial_\gamma \psi)(\partial_\gamma \psi) \delta_{\alpha \beta},
\end{equation}
\begin{equation} 
\label{eq:PTCST}
p^{(2)}_{\alpha \beta} = \psi \partial_\alpha \partial_\beta \psi - (\psi \partial_\gamma \partial_\gamma \psi) \delta_{\alpha \beta}.
\end{equation}
\end{subequations}

The term $p_{\alpha \beta}^{(1)}$ affects directly the value of the parameter $\epsilon$
in Eq.~(\ref{eq:PGLDR}), influencing the shape of the saturation curve. On the other
hand the pressure $p_{\alpha \beta}^{(2)}$ is related with surface tension, because the
first term of the right hand side of Eq.~(\ref{eq:PTCST}) is anisotropic. Note that this
term also do not affect the $\epsilon$ parameter, thus not changing the density relation
for the planar interface. In such way, by introducing these terms, the deficiencies of 
Shan-Chen method can be corrected. These pressure terms can be converted in equivalent 
forces in the macroscopic governing equations using the relations:  
\begin{subequations}
\begin{equation}
\label{eq:CCFT} 
F_{\alpha}^{(1)} = - \partial_{\beta} p_{\alpha \beta}^{(1)} = - 2 (\partial_{\beta} \psi)(\partial_{\alpha} \partial_{\beta} \psi),
\end{equation}
\begin{equation}
\label{eq:STFT} 
F_{\alpha}^{(2)} = - \partial_{\beta} p_{\alpha \beta}^{(2)} = - \partial_{\beta} \left[ \psi \partial_{\alpha} \partial_{\beta} \psi - (\psi \partial_{\gamma} \partial_{\gamma} \psi) \delta_{\alpha \beta} \right].
\end{equation}
\end{subequations}

Using the tensor identity, Eq.~(\ref{eq:TI}), it is possible to rewrite Eq.~(\ref{eq:STFT}) to
the following form:
\begin{equation}
\label{eq:SSTFT} 
\begin{aligned}
F_{\alpha}^{(2)} =& - \partial_{\beta} \left[ (\partial_{\gamma} \psi) (\partial_{\gamma} \psi) \delta_{\alpha \beta}
- (\partial_{\alpha} \psi) (\partial_{\beta} \psi) \right] \\
=&~(\partial_{\alpha} \psi)(\partial_{\beta} \partial_{\beta} \psi) - (\partial_{\beta} \psi)(\partial_{\alpha} \partial_{\beta} \psi). 
\end{aligned}
\end{equation}

Now that it was obtained explicit expressions for the interaction forces, it is necessary
to replace the spatial derivatives for numerical approximations. This can be done with
the moments defined by Eqs.~(\ref{eq:ZOMED}), (\ref{eq:FOMED}) and (\ref{eq:SOMED}).
Truncating the series in its first term and replacing the derivatives, the following 
expressions are obtained: 
\begin{subequations}
\begin{equation} 
\label{eq:DFTCCC}
F^{(1)}_{\alpha} = - 2 \frac{M^{1}_{\beta}}{c_s^2} \frac{M^{2}_{\alpha \beta}}{c_s^4},
\end{equation}
\begin{equation} 
\label{eq:DFTCST}
F^{(2)}_{\alpha} = \frac{M^{1}_{\alpha}}{c_s^2} \frac{M^{2}_{\beta \beta}}{c_s^4} - 
\frac{M^{1}_{\beta}}{c_s^2} \frac{M^{2}_{\alpha \beta}}{c_s^4}.
\end{equation}
\end{subequations}

Comparing Eq.~(\ref{eq:ZOMED}) and (\ref{eq:SOMED}) it can be concluded that another 
option is to use $M^{2}_{\beta \beta} = 2 c_s^2 (M^0 - \psi)$. Eq.~(\ref{eq:DFTCCC}) and 
(\ref{eq:DFTCST}) are very useful and can be used to improve the Shan-Chen pseudopotential
method to achieve thermodynamic consistency and adjustable surface tension. Based
on this finding, the force term showed in Eq.~(\ref{eq:IFT}) is proposed:
\begin{equation} 
\label{eq:IFT}
F_\alpha = F^{SC}_{\alpha} - \frac{3}{4} \epsilon c_s^2 G F^{(1)}_{\alpha} 
+ \left( \sigma - 1 \right) c_s^2 G F^{(2)}_{\alpha}.
\end{equation}

This force represents the general force proposed in Sec.~\ref{sec:GIPFP}. The tensor $A_{ij}$ 
is derived as follow. Expressing the fitting potential function as
$\psi(\bm{x})=\psi(\bm{x}+\bm{c}_0)$ with $\bm{c}_0=0$, one can write:
\begin{eqnarray} 
\psi(\bm{x}+\bm{c}_0) = \sum_j\psi(\bm{x}+\bm{c}_j)\delta_{j0}.
\end{eqnarray}
The Shan-Chen force, Eq.~(\ref{eq:SCPTWDE}), can be rewritten as:
\begin{eqnarray}
F_{\alpha}^{SC} = \sum_i \sum_j \left[ -\frac{2G}{c_s^2} w_i \delta_{j0} \right] 
\psi(\bm{x}+\bm{c}_i) \psi(\bm{x}+\bm{c}_j).
\end{eqnarray}

Combining Eqs.~(\ref{eq:FOMED}) and (\ref{eq:SOMED}) with Eq.~(\ref{eq:DFTCCC}) results: 
\begin{eqnarray} 
F_{\alpha}^{(1)} = &&~ -2 \left[ \sum_i \frac{w_i}{c_s^2} c_{i\beta} \psi(\bm{x}+\bm{c}_i) \right] \nonumber \\
&& \times \left[ \sum_j \frac{w_j}{c_s^4} (c_{j\alpha}c_{j\beta}-c_s^2 \delta_{\alpha \beta})
\psi(\bm{x}+\bm{c}_j) \right], \nonumber \\
F_{\alpha}^{(1)} = &&~ \sum_i \sum_j \left[ -2 \frac{w_i}{c_s^2} \frac{w_j}{c_s^4}
c_{i\beta}(c_{j\alpha}c_{j\beta}-c_s^2\delta_{\alpha\beta}) \right] \nonumber \\
&& \times~ \psi(\bm{x}+\bm{c}_i) \psi(\bm{x}+\bm{c}_j).
\end{eqnarray}
Now noting that the second term of the right side of Eq.~(\ref{eq:DFTCST}) is equal to 
$F_{\alpha}^{(1)}/2$ and that the first term can be calculated with the help of 
Eqs.~(\ref{eq:FOMED}) and (\ref{eq:SOMED}) as follows:
\begin{eqnarray} 
\frac{M_{\alpha}^{1}}{c_s^2} \frac{M_{\beta\beta}^{2}}{c_s^4} =
\left[ \sum_i \frac{w_i}{c_s^2} c_{i\alpha} \psi(\bm{x}+\bm{c}_i) \right] \nonumber \\
\times \left[ \sum_j \frac{w_j}{c_s^4} (c_{j\beta}c_{j\beta}-c_s^2\delta_{\beta\beta}) \right], \nonumber \\
= \sum_i \sum_j \left[ \frac{w_i}{c_s^2} \frac{w_j}{c_s^4}
c_{i\alpha} (c_{j\beta}c_{j\beta}-c_s^2\delta_{\beta\beta}) \right] \nonumber \\ 
\times~\psi(\bm{x}+\bm{c}_i) \psi(\bm{x}+\bm{c}_j).
\end{eqnarray}

The term $F_{\alpha}^{(2)}$ is formulated as:
\begin{eqnarray}
F_{\alpha}^{(2)} = \sum_i \sum_j \frac{w_i}{c_s^2} \frac{w_j}{c_s^4}
\bigg[ c_{i\alpha} (c_{j\beta}c_{j\beta}-c_s^2\delta_{\beta\beta}) \nonumber \\
- c_{i\beta}(c_{j\alpha}c_{j\beta}-c_s^2\delta_{\alpha\beta}) \bigg]
\psi(\bm{x}+\bm{c}_i) \psi(\bm{x}+\bm{c}_j),
\end{eqnarray}
substituting the above relations into the Eq.~(\ref{eq:IFT}),
the $A_{ij}$ tensor from Eq.~(\ref{eq:Ansatz}) can be determined:
\begin{eqnarray} 
A_{ij} = -\frac{2G}{c_s^2} w_i \delta_{j0} \nonumber \\
+ \left[ \frac{3}{2} \epsilon - ( \sigma - 1 ) \right] c_s^2 G 
\frac{w_i}{c_s^2} \frac{w_j}{c_s^4} c_{i\beta} (c_{j\alpha}c_{j\beta}-c_s^2\delta_{\alpha\beta}) \nonumber \\
+ ( \sigma - 1 ) c_s^2 G 
\frac{w_i}{c_s^2} \frac{w_j}{c_s^4} c_{i\alpha} (c_{j\beta}c_{j\beta}-c_s^2\delta_{\beta\beta}).
\end{eqnarray}

When the above force is incorporated into the lattice Boltzmann equation using the
\citeauthor{guo2002discrete} force scheme, it results in the following pressure tensor in the momentum
conservation equation:
\begin{eqnarray} 
\label{eq:PTIF}
p_{\alpha \beta} = p^{SC}_{\alpha \beta} - \frac{3}{4} \epsilon c_s^2 G p^{(1)}_{\alpha \beta} 
+ \left( \sigma - 1 \right) c_s^2 G p^{(2)}_{\alpha \beta},
\end{eqnarray}
using Eqs.~(\ref{eq:CFPTSF}), (\ref{eq:PTCCC}) and (\ref{eq:PTCST}) in the above relation it is 
obtained the final expression:
\begin{eqnarray} 
\label{eq:PTIFT}
p_{\alpha \beta} = &&~ \bigg( c_s^2 \rho + G \psi^2 - \frac{3}{4} \epsilon c_s^2 G (\partial_\gamma \psi)(\partial_\gamma \psi) \nonumber \\
&&~ + \left( \frac{3}{2} - \sigma \right)  c_s^2 G \psi \partial_{\gamma} \partial_{\gamma} \psi \bigg) \delta_{\alpha \beta}
\nonumber \\
&&~ + \sigma c_s^2 G \psi \partial_{\alpha} \partial_{\beta} \psi.
\end{eqnarray}

The force was written in such a way that the parameter $\epsilon$, Eq.~(\ref{eq:CTC}), of the consistency condition
appears explicitly. This way, it is easy to adjust the coexistence curve and then control
the surface tension through the coefficient $\sigma$ of the anisotropic term of the
pressure tensor Eq.~(\ref{eq:PTIFT}). 




\section{Numerical Simulations} 
\label{sec:NS}


\subsection{Static droplet and coexistence curve} 
\label{sec:SDCC}

The first aspect of the presented model that is tested is the ability of control
the coexistence curve of the pseudopotential method by the choice of the parameter
$\epsilon$. Numerical simulations were performed using Carnahan-Starling (C-S) equation 
of state: 
\begin{eqnarray} 
\label{eq:CSEOS}
P_{EOS} = k \bigg[ c \rho T \frac{1 + b\rho + (b\rho)^2 - (b\rho)^3}{(1-b\rho)^3} - a \rho^2 \bigg],
\end{eqnarray}
the parameters were chosen to be $a=3.852462257$, $b=0.1304438842$ and $c=2.785855166$ which 
are the same values used in reference \cite{kupershtokh2009equations}. These authors
introduced the scaling factor $k$ in Eq.~(\ref{eq:CSEOS}), which can also be used to increase
the stability of the pseudopotential method \cite{hu2013equations}. This parameter is set as $k=0.01$.
Following \cite{huang2011forcing}, the computational domain is given by a mesh of 200 $\times$ 200 nodes 
with periodic boundary condition. A liquid droplet is initialized in the center of the domain using 
the function:
\begin{eqnarray} 
\label{eq:ISD}
\rho(x,y) = \frac{\rho_{l}+\rho_{g}}{2} - \frac{\rho_l-\rho_g}{2} \text{tanh} 
\bigg[ \frac{2 (R-R_0)}{W} \bigg],
\end{eqnarray}
where $W=5$ and $R=\sqrt{(x-x_0)^2+(y-y_0)^2}$, with $(x_0,y_0)$ being the central position
of the computational domain. For a specific temperature, the values of $\rho_g$ and $\rho_l$
are initialized as the saturation densities obtained with Maxwell equal area rule. 
The velocity was set as zero everywhere.
With the initialization of macroscopic fields, the equilibrium distribution function, 
Eq.~(\ref{eq:SEDF}), is determined in each lattice and the particle distribution function 
is set as equal to the equilibrium function. The lattice Boltzmann equation was solved using
the BGK collision operator, Eq.~(\ref{eq:BGKCO}), with $\tau = 0.8$. Simulations were carried until the following convergence criteria has being obeyed:
\begin{equation} 
\label{eq:SSC}
\frac{ \sum \mid [\rho(t)-\rho(t-100)] \mid }{ \sum \mid \rho(t) \mid } < 10^{-6}.
\end{equation}

\begin{figure}
	\includegraphics[width=80mm]{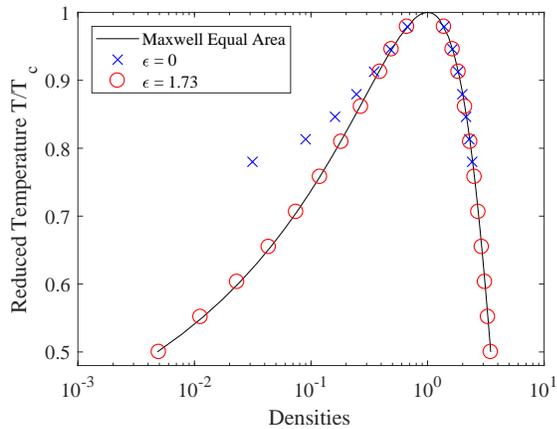}
	\caption{Coexistence densities curves for planar interface. Comparison between 
	results obtained with the lattice Boltzmann simulations for two values of 
	the $\epsilon$ parameter and the results obtained analytically with the Maxwell
	equal area rule.}
	\label{fig:SDCC}
\end{figure}

Setting $G=-1$, $\epsilon=0$, $\sigma=1$ in Eq.~(\ref{eq:IFT}), simulations were performed
for different temperatures. One can notice that using these parameters is equivalent to
use the original Shan-Chen force, Eq.~(\ref{eq:SCF}). Then, the value of $\epsilon$ is
adjusted until the saturation curve of the pseudopotential method matches with the one given
by the equal area rule. The value $\epsilon=1.73$ was found to provide this adjustment. 
Results can be seen in Fig.~(\ref{fig:SDCC}).

\subsection{Young-Laplace Test} 
\label{sec:YLT}

In order to evaluate the influence of the parameter $\sigma$ in the surface tension,
further static droplet simulations were performed. 
The numerical procedure is similar to the ones
used to produce the coexistence curve with the difference that all tests are conducted
with a fixed temperature of $T_r=0.8$ and fixed $\epsilon$ of value 1.73. Then the parameter
$\sigma$ is specified and when the droplet reaches an equilibrium stage, the surface tension 
is measured using: 
\begin{equation} 
\label{eq:YLR}
\Delta P = \gamma \bigg( \frac{1}{R_1} + \frac{1}{R_2} \bigg),
\end{equation}
also known as the Young-Laplace relation. Fixing $\sigma$, simulations
are performed for different radius and it is expected that the surface tension to remain 
constant. In the end, all this procedure is repeated for different values of $\sigma$. 
In Eq.~(\ref{eq:YLR}), $\gamma$ is the surface tension, $\Delta P$ is the pressure difference across
the interface. The parameters $R_1$ and $R_2$ are the radius of curvature of the
interface. In the present case of a planar (2 dimension) droplet, there is only
one radius of curvature equal to the radius of the droplet. In order to measure the
radius, it was defined that the interface of the droplet was located in the region that
the density is equal to $\rho_m=(\rho_l+\rho_g)/2$. 

\begin{figure}
\includegraphics[width=80mm]{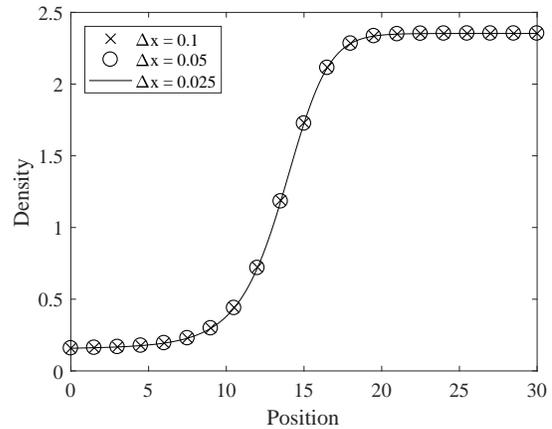}
\caption{Theoretical density profile for a planar interface case. It was adopted $\epsilon=1.73$ and a reduced temperature of $T_r=0.8$. 
Boundary conditions were $\rho_g=0.1580$ and $\rho_l=2.3530$.}
\label{fig:TDP}
\end{figure}

In order to obtain a comparison for the lattice Boltzmann simulation results,
the surface tension of the planar interface case was computed using the procedure described in the end of Sec. \ref{sec:PMPC}.
Eq.~(\ref{eq:PTNNPI}) was solved numerically to obtain the density profile.
One should note that this equation depends only on the values of $A_1$ and $A_2+A_3$, which are given by Eq.~(\ref{eq:PTIFT}), being $A_1=-3 \epsilon c_s^2/4$ and $A_2 + A_3=3c_s^2/2$. 
In this way the density profile does not depend on the $\sigma$ parameter, which influences only the surface tension value by the
coefficient $A_3$ in Eq.~(\ref{eq:STPI}), given by $A_3=\sigma c_s^2$.
The boundary conditions used are the phase densities. For a reduced temperature
$T_r=0.8$ and $\epsilon=1.73$, Eq.~(\ref{eq:PGLDR}) provide $\rho_{g}\approx0.1580$ and $\rho_{l}\approx2.3530$
as the vapor and liquid densities, respectively.
A spatial domain of length $L=30$ was used and the differential equation was solved using three different mesh sizes $\Delta x=0.1$, $\Delta x=0.05$ and $\Delta x=0.025$. 
The density profile is shown in Fig.~(\ref{fig:TDP}). It can be observed convergence of results since the profiles are very close even with the mesh refinement. 

\begin{figure}
\includegraphics[width=80mm]{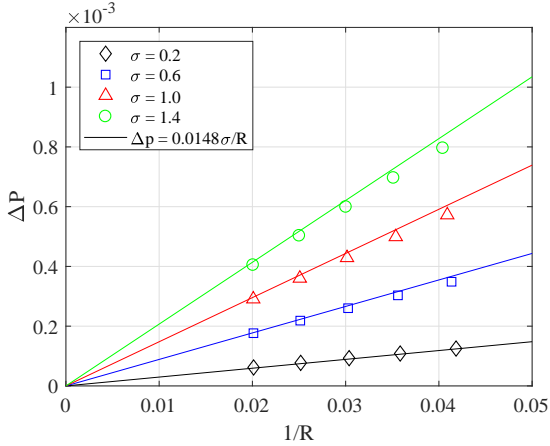}
\caption{Young-Laplace tests for static droplet with $\sigma$ varying from 0.2 to 1.4 are performed for $\epsilon=1.73$ and 
reduced temperature $T_r=0.8$. The solid lines represents 
the theoretical surface tension for a planar interface $\gamma_{pi}(\sigma)$.}
\label{fig:YLT}
\end{figure}

After that the surface tension was computed using Eq.~(\ref{eq:STPI}).
It was obtained that the surface tension for a planar interface $\gamma_{pi}$ is given by 
the expression $\gamma_{pi}\approx0.0148\sigma$, for the specified conditions. 
A comparison between this results with the Young-Laplace test can be seen in Fig.~(\ref{fig:YLT}).
It is expected a small difference between the surface tension
values obtained with the droplet and the planar interface tests
because for the second case, the density profile is obtained 
considering that the pressure is constant along the normal direction to the interface. This is not true for the 
static droplet case. However, for large droplet radius, one may expect a
better agreement in results, since the interface curvature tends to zero, which approximates the case to a planar 
interface problem. And this is exactly the behavior observed in Fig.~(\ref{fig:YLT}). For a radius of $50$ lattice sites,
which corresponds to $1/R=0.02$, the results of both cases were
very close. It was also observed that the method succeeds in controlling the surface tension by adjusting the parameter $\sigma$.

The force term, Eq.~(\ref{eq:IFT}), was devised in such a way that the surface tension
could be adjusted without affecting the coexistence densities. In order to test this property,
further tests were performed. The static droplet was simulated in a similar way as previous examples,
but in this case, it was set $T_r=0.8$, $\epsilon=1.73$, $R_0=50$ (initial radius of the droplet) 
and only $\sigma$ was varied. For each test, the surface tension and the densities of the
liquid and gas phases were measured. The results can be seen in Table~(\ref{tab:DVT}).
A comparison was carried out between the surface tension obtained
by simulations ($\gamma$) with the planar interface theoretical value ($\gamma_{pi}$),
for the same reduced temperature and $\epsilon$ parameter. Also, the phase densities results were compared
with the ones obtained by the Maxwell equal area rule, which are given by $\rho_{gm}=0.1665$ and $\rho_{lm}=2.3550$ for the vapor and liquid phase, respectively. 
On Table~(\ref{tab:DVT}), it is observed that the surface tension can be widely varied without affecting significantly the phase densities. 
\begin{table*}
\caption{Comparison between the variation of the surface tension with the variation of the 
	liquid and vapor densities, obtained with the adjust of the parameter $\sigma$
	for a droplet of radius $R=50$ modelled by the C-S equation of state with a reduced
	temperature $T_r=0.8$. It was also presented the theoretical surface tension value for
	a planar interface $\gamma_{pi}$ and a comparison between the phase densities obtained
	by simulations with the vapor and liquid densities consistent with the Maxwell equal area
	rule, given by $\rho_{gm}=0.1665$ and $\rho_{lm}=2.3550$.} \label{tab:DVT}
\begin{ruledtabular}
\begin{tabular}{cccccccc}
$\sigma$ & $\gamma$ & $\gamma_{pi}$ & $100\cdot\gamma/\gamma_{pi}$ & $\rho_g$ & $100\cdot\rho_g/\rho_{gm}$ & $\rho_l$ & $100\cdot\rho_l/\rho_{lm}$ \\ \hline
4 & 0.0603 & 0.0592 & 101.86 & 0.1595 & 95.80 & 2.3725 & 100.74 \\
2 & 0.0290 & 0.0296 & 97.97 & 0.1658 & 99.58 & 2.3644 & 100.40 \\
1 & 0.0145 & 0.0148 & 97.97 & 0.1688 & 101.40 & 2.3603 & 100.23 \\
1/2 & 0.0074 & 0.0074 & 100.00  & 0.1704 & 102.34 & 2.3583 & 100.14 \\
1/4 & 0.0039 & 0.0037 & 105.41 & 0.1711 & 102.76 & 2.3573 & 100.10 \\
1/8 & 0.0020 & 0.00185 & 108.11 & 0.1715 & 103.00 & 2.3568 & 100.08 \\
\end{tabular}
\end{ruledtabular}
\end{table*}



\subsection{Droplet Oscillation} 
\label{sec:DO}

The next case is a dynamic test. It consists in a elliptic droplet oscillating in a vapor
medium. Here, the C-S equation of state was used again. 
Two simulations were conducted, for the reduced temperatures $T_r=0.6$ and $T_r=0.7$.
The surface tension values and the phase densities for these reduced temperatures
can be seen in Table~(\ref{tab:PRCases}). 
The Young-Laplace test was applied to obtain the values of the surface tension.
\begin{table}[htbp]
\caption{\label{tab:PRCases} 
		Saturation densities and surface tension obtained through static droplet test for the 
		reduced temperatures $T_r=0.6$ and $T_r=0.7$ using the Carnahan-Starling (C-S) equation
		of state.
}
\begin{ruledtabular}
\begin{tabular}{cccc}
\textrm{$T_r$}&
\textrm{$\rho_g$}&
\textrm{$\rho_l$}&
\textrm{$\gamma$}\\
\colrule
0.6 & 0.0224 & 3.1192 & 0.0461\\
0.7 & 0.0700 & 2.7504 & 0.0267\\
\end{tabular}
\end{ruledtabular}
\end{table}
\begin{figure}
	\includegraphics[width=80mm]{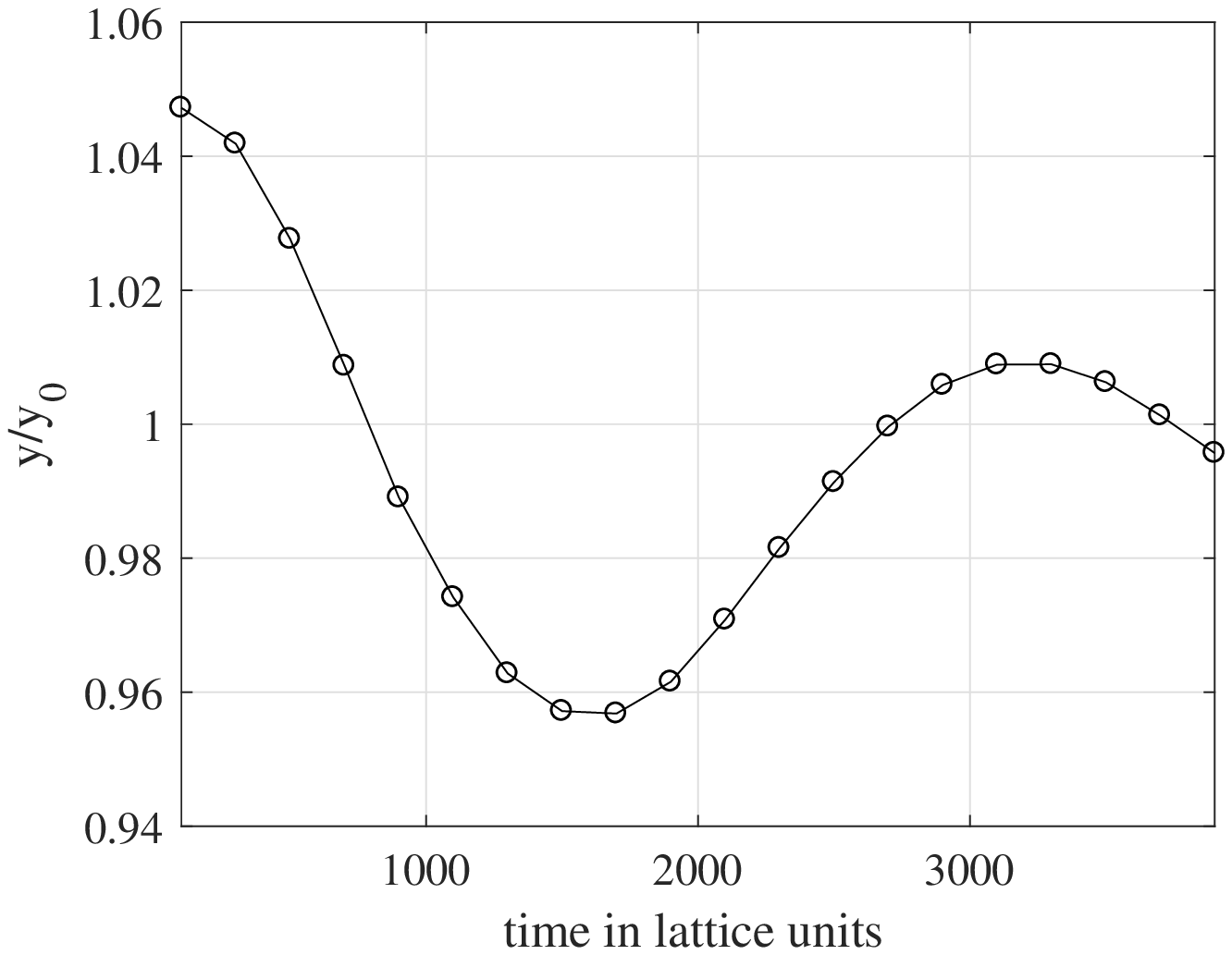}
	\caption{Oscillation of an elliptic droplet for a fluid modelled by 
	            the C-S equation of state with reduced temperature Tr = 0.6.}
	\label{fig:DO1} 
\end{figure}
\begin{figure}
	\includegraphics[width=80mm]{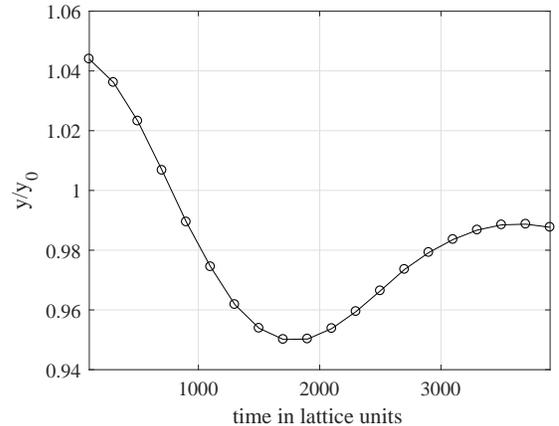}
	\caption{Oscillation of an elliptic droplet for a fluid modelled by 
	            the C-S equation of state with reduced temperature Tr = 0.7.}
	\label{fig:DO2} 
\end{figure}

It is desired to initialize an elliptic profile of major radius $R_{max}=30$ and minor radius
$R_{min}=27$ in a $200 \times 200$ grid. As the pseudopotential method is a diffuse interface technique, 
a diffuse profile is initialized using Eq.~(\ref{eq:ISD}). But now $R_o$ is a function of space coordinates
$R_0=R_0(x,y)$ and it is given by the following relations:
\begin{subequations}
\begin{equation} 
\label{eq:REL}
R_0(\theta) = \frac{R_{min}}{ \sqrt{ 1-(e\cos(\theta))^2 } },
\end{equation}
\begin{equation} 
\theta(x,y) = \arctan \left( \frac{y-y_0}{x-x_0} \right),
\end{equation}
\begin{equation} 
e = \sqrt{1- \left( \frac{R_{min}}{R_{max}} \right)^2},
\end{equation}
\end{subequations}
with $(x_0,y_0)$ being the central position of the computational domain. The initial 
distribution function field is initialized equal to the equilibrium function 
$f_i(t=0,\bm{x})=f_i^{eq}(t=0,\bm{x})$. It is clear that the initial state is not in equilibrium,
so it is expected some error due to the chosen initialization procedure. To solve this case,
it is used the lattice Boltzmann equation with the Gram-Shmidt based MRT collision operator, Eq.~(\ref{eq:MRTCO}).
This option is based on the fact that this collision term is more stable at low viscosity, which
is necessary in a dynamic test as viscosity dissipates perturbations rapidly. The force 
scheme used is given by Eqs.~(\ref{eq:GFS}) and (\ref{eq:GFSMRT}). The relaxation matrix 
(more details in Appendix~\ref{sec:MRT}) used is given by: 
\begin{eqnarray} 
\label{eq:RM}
\bm{\Lambda} = diag\left( 1,1,1,1,1,1,1,\tau^{-1},\tau^{-1} \right),
\end{eqnarray}
here, it was used $\tau=0.65$ which results in a kinematic viscosity $\nu=(\tau-0.5)/3=0.05$.
The droplet oscillation period is given analytically, according to \cite{lamb1932hydrodynamics},
by the relation:
\begin{eqnarray} 
\label{eq:DOP} 
T_a = 2 \pi \left[ n(n^2-1) \frac{\gamma}{\rho_l R_m^3} \right]^{-\frac{1}{2}},
\end{eqnarray}
where $R_m=\sqrt{R_{max}R_{min}}$ and $n=2$ for an initial elliptic shape \cite{li2013lattice,mukherjee2007pressure}.
The analytical result for $T_r=0.6$ is $T_a\approx3204$. The simulation is conducted for $4000$ time steps. 
The distance between the right extremity of the ellipse to its center is measured at each $100$ time steps. 
The results are shown on Fig.~(\ref{fig:DO1}). 
The numerical period of oscillation obtained is $T_n=3200$ which represents an absolute relative error 
of $0.1\%$ of the analytical solution. 

For the case with reduced temperature $T_r=0.7$ the droplet has a thicker interface width in
comparison with the case for $T_r=0.6$. In this way, it is expected a larger deviation in the 
solution. The analytical result using information from 
Table~(\ref{tab:PRCases}) is $T_a\approx3953$. Again, the distance between the right extremity of the 
ellipse to its center is measured at each $100$ time steps. The numerical period of oscillation is $T_n = 3600$
which represents an absolute relative error of $9\%$ of the analytical solution. Results are shown in Fig.~(\ref{fig:DO2}).




\section{Conclusion} 
\label{sec:Conclusion}

In the present work, an interaction force able to control the liquid-gas density ratio
and the surface tension in the pseudopotential LBM was devised. 
First, the pressure tensor was written in a generic form and the role of each term was
analyzed. Attention was paid to the property that different pressure tensors can result in the same divergence, reducing the number of terms necessary to describe
the pressure tensor. 
After, the \citeauthor{shan1993lattice} model was studied by means of an 
equivalent pressure tensor including the third order spatial discretizetion errors caused by the Guo forcing scheme. 

Later, it was presented finite difference approximations for the terms that constitute
the pressure tensor. This approximations were written in the same notation as the 
moments of the distribution function. 
To devise the new interaction force, suitable terms of the generic pressure tensor were
selected to complement the \citeauthor{shan1993lattice} model. Then it was derived
an external force field able to replicate the effects of this pressure tensor terms in 
the conservation equations. This force field was converted into a numerical scheme using
the finite difference approximations presented in Sec.~\ref{sec:MOTED}. The result
is a numerical force to be implemented into the LBM with the Guo forcing scheme. 

Numerical simulations of a static droplet showed the ability of the method in control
the liquid-gas density ratio and surface tension. Also, good results with dynamic tests
were obtained. The proposed numerical scheme is versatile as the force was tested with
BGK and MRT collision operator with no change in the procedure to calculate the external
force. The new feature of this force is that it permits the control of these multiphase properties considering only nearest-neighbor interactions, which provides computational efficiency in comparison with current interaction forces available in the literature. 


\begin{acknowledgments}
The authors acknowledge the support received from CAPES (Coordination for the Improvement of Higher Education Personnel, Finance Code 001), from CNPq (National Council for Scientific and Technological Development, process 304972/2017-7) and FAPESP (S\~ao Paulo Foundation for Research Support, 2016/09509-1 and 2018/09041-5), for developing research that have contributed to this study.
\end{acknowledgments}

\appendix


\section{MRT Matrix} 
\label{sec:MRT}

The MRT collision operator was presented in Eq.~(\ref{eq:MRTCO}). This operator depends
on the matrix $\bm{M}$ that converts the distribution functions into a set of linear independent
moments. In this work it is used $\bm{M}$ obtained by a Gram-Schmidt procedure 
\cite{kruger2017lattice} which is given by the following relation:
\begin{equation} 
\label{eq:MatrixMRT}
\bm{M} = 
 	\begin{pmatrix}
 		1 & 1 & 1 & 1 & 1 & 1 & 1 & 1 & 1  \\
 		-4 & -1 & -1 & -1 & -1 & 2 & 2 & 2 & 2 \\ 
 		4 & -2 & -2 & -2 & -2 & 1 & 1 & 1 & 1 \\
 		0 & 1 & 0 & -1 & 0 & 1 & -1 & -1 & 1 \\
		0 & -2 & 0 & 2 & 0 & 1 & -1 & -1 & 1 \\
 		0 & 0 & 1 & 0 & -1 & 1 & 1 & -1 & -1 \\
 		0 & 0 & -2 & 0 & 2 & 1 & 1 & -1 & -1 \\
 		0 & 1 & -1 & 1 & -1 & 0 & 0 & 0 & 0 \\
 		0 & 0 & 0 & 0 & 0 & 1 & -1 & 1 & -1 \\
 		\end{pmatrix},
\end{equation}
while the relaxation matrix $\bm{\Lambda}$ can be written as: 
\begin{equation} 
\label{eq:RelaxationMatrix}
\bm{\Lambda} = diag\left( \tau_{\rho}^{-1},\tau_{e}^{-1},\tau_{\varsigma}^{-1},
\tau_{j}^{-1},\tau_{q}^{-1},\tau_{j}^{-1},\tau_{q}^{-1},
\tau_{\nu}^{-1},\tau_{\nu}^{-1} \right).
\end{equation}
The relaxation time $\tau_{\nu}$ controls the fluid viscosity
by the relation $\mu=\rho c_s^2(\tau_{\nu}-0.5)$.
A set of moments of the equilibrium distribution function $\bm{m}^{eq}$ is obtained 
by multiplying the matrix $M$, Eq.~(\ref{eq:MatrixMRT}), by the equilibrium distribution function
vector, $\bm{f}^{eq}$, with components $f_i=[\bm{f}^{eq}]_i$ given by Eq.~(\ref{eq:SEDF}):
\begin{equation}
\bm{m}^{eq} = \bm{M}\bm{f}^{eq} = 
 	\begin{pmatrix}
 		\rho \\
 		-2 \rho + 3 \rho |\bm{u}|^2 \\ 
 		\rho - 3 \rho |\bm{u}|^2 \\
 		\rho u_x \\
		-\rho u_x \\
 		\rho u_y \\
 		-\rho u_y \\
 		\rho \left( u_x^2 - u_y^2 \right) \\
 		\rho u_x u_y \\
 		\end{pmatrix},
\end{equation}
and the force scheme in the moment space $\bm{\overline{S}}=\bm{M}\bm{S}$ can be
written as:
\begin{equation}
\bm{\overline{S}} = \bm{M}\bm{S} = 
 	\begin{pmatrix}
 		0 \\
 		6 \left( u_x F_x + u_y F_y \right) \\ 
 		-6 \left( u_x F_x + u_y F_y \right) \\
 		F_x \\
		-F_x \\
 		F_y \\
 		-F_y \\
 		2 \left( u_x F_x - u_y F_y \right) \\
 		u_x F_y + u_y F_x  \\
 		\end{pmatrix}.
\end{equation}



\section{Tensor Identity} 
\label{sec:TI}

In Section \ref{sec:PMPC}, it was discussed how the divergence of different pressure 
tensors can lead to the same result. And an identity was provided by Eq.~(\ref{eq:TI}).
In this appendix the given identity will be proven. The left-hand side of
Eq.~(\ref{eq:TI}) is represented by Eq.~(\ref{eq:LHSone}) and the right-hand side by Eq.~(\ref{eq:RHSone}):
\begin{subequations}
\begin{equation} 
\label{eq:LHSone} 
L_{\alpha} = \partial_{\beta} \big[ \psi \partial_{\alpha} \partial_{\beta} \psi - \big( \psi \partial_{\gamma} \partial_{\gamma} \psi \big) \delta_{\alpha \beta} \big],
\end{equation} 	 
\begin{equation} 
\label{eq:RHSone} 
R_{\alpha} = \partial_{\beta} \big[ (\partial_{\gamma} \psi)(\partial_{\gamma} \psi)
\delta_{\alpha \beta} - (\partial_{\alpha} \psi)(\partial_{\beta} \psi) \big],
\end{equation} 	 
\end{subequations}
applying the sum rule for derivative and the fact that $\partial_{\beta}(a)\delta_{\alpha \beta} = \partial_{\alpha}a$, where $a$ is a scalar:
\begin{subequations}
\begin{equation} 
\label{eq:LHStwo} 
L_{\alpha} = \partial_{\beta} (\psi \partial_{\alpha} \partial_{\beta} \psi)
- \partial_{\alpha} \big( \psi \partial_{\gamma} \partial_{\gamma} \psi \big),
\end{equation} 
\begin{equation} 
\label{eq:RHStwo} 
R_{\alpha} = \partial_{\alpha} \big[ (\partial_{\gamma} \psi)(\partial_{\gamma} \psi) \big]
- \partial_{\beta} \big[ (\partial_{\alpha} \psi)(\partial_{\beta} \psi) \big].
\end{equation} 
\end{subequations}

Now, applying the product rule for derivative,
Eqs.~(\ref{eq:LHStwo}) and (\ref{eq:RHStwo}) can be rewritten in the following way:
\begin{subequations}
\begin{align} 
\label{eq:LHSthree} 
L_{\alpha} = &&~ (\partial_{\beta} \psi)(\partial_{\alpha} \partial_{\beta} \psi)
+ \psi \partial_{\alpha} \partial_{\beta} \partial_{\beta} \psi \nonumber \\
&& - (\partial_{\alpha} \psi)(\partial_{\gamma} \partial_{\gamma} \psi)
- \psi \partial_{\alpha} \partial_{\gamma} \partial_{\gamma} \psi,
\end{align} 	 
\begin{align} 
\label{eq:RHSthree} 
R_{\alpha} = &&~ 2(\partial_{\gamma} \psi)(\partial_{\alpha} \partial_{\gamma} \psi)
- (\partial_{\alpha} \psi)(\partial_{\beta} \partial_{\beta} \psi) \nonumber \\
&& - (\partial_{\beta} \psi)(\partial_{\alpha} \partial_{\beta} \psi),
\end{align} 	 
\end{subequations}
the dummy index $\gamma$ can be replaced without affecting the results, so choosing $\beta$
in its place:
\begin{subequations}
\begin{equation} 
\label{eq:LHSfour} 
L_{\alpha} = (\partial_{\beta} \psi)(\partial_{\alpha} \partial_{\beta} \psi)
- (\partial_{\alpha} \psi) (\partial_{\beta} \partial_{\beta} \psi),
\end{equation} 	 
\begin{equation} 
\label{eq:RHSfour} 
R_{\alpha} = (\partial_{\beta} \psi)(\partial_{\alpha} \partial_{\beta} \psi)
- (\partial_{\alpha} \psi) (\partial_{\beta} \partial_{\beta} \psi).
\end{equation} 	
\end{subequations}

Now, the equality is proven.




\bibliography{Main}

\end{document}